\documentclass{aleph-revista}

\usepackage{tikz}           % Para generar gráficos 
\usepackage{aleph-comandos} % Comandos para facilitar escritura matemática
\usepackage{multicol}       % Para utilizar multicolumnas
\usepackage{arydshln} % Linha tracejada em tabela
\usepackage{multirow}
\usepackage{pdfpages}
\usepackage{subcaption}

\volumen{5}
\numero{1}
\fechapubli{2021}
\periodouno{Abril}
\periododos{Julio}
\titulo{Proposal of representative portfolios for federal roadway bridges in Northeastern Brazil}

\autor{%
    Gustavo Henrique Ferreira Cavalcante\textsuperscript{1 \faEnvelopeO}\\ 
    Eduardo Marques Vieira Pereira\textsuperscript{1}\\ 
    Isabela Durci Rodrigues\textsuperscript{2}\\
    Luiz Carlos Marcos Vieira Júnior\textsuperscript{1}\\ 
    Jamie Ellen Padgett\textsuperscript{3}\\ 
    Gustavo Henrique Siqueira\textsuperscript{1}\\ 
}

\institucion{
\textsuperscript{1}%
    Universidade Estadual de Campinas – Unicamp, Faculdade de Engenharia Civil, Arquitetura e Urbanismo, Departamento de Estruturas, Campinas, São Paulo, Brasil\\
\textsuperscript{2}%
    Universidade de São Paulo – USP, Escola de Engenharia de São Carlos, Departamento de Estruturas, São Carlos, São Paulo, Brasil\\
\textsuperscript{3}%
    Rice University – Department of Civil and Environmental Engineering, Houston, Texas, United States
}

\correo{ghenriquefc@hotmail.com}

\abstract{
    This paper presents a statistical analysis of federal highway bridges commonly found in Northeastern Brazil to develop a portfolio, or statistically representative characterization of bridges across the region. A detailed study of bridges under the supervision of the National Department of Infrastructure and Transportation is conducted and four representative bridge classes are defined: two of them consist of single-span bridges and the others are multi-span continuous bridges with non-integral or no abutments and different bridge decks. Discrete and continuous distributions describe random variables to consider their variability in the analyses. However, some parameters are defined as function of the random variables, since a strong correlation is observed. Future bridge assessment studies should use the geometries of this bridge portfolio to evaluate the regional impacts due to natural hazards, or furnish models for updating, among other applications.
}

\begin{document}
%%%%%%%%%%%%%%%%%%%%%%%%%%%%%%%%%%%%%%%%
%% Encabezado
%%%%%%%%%%%%%%%%%%%%%%%%%%%%%%%%%%%%%%%%
\membrete

\newpage
%%%%%%%%%%%%%%%%%%%%%%%%%%%%%%%%%%%%%%%%
\section{Introduction}
%%%%%%%%%%%%%%%%%%%%%%%%%%%%%%%%%%%%%%%%
Bridges have fundamental importance in the economic and social development of cities. In this context, it is essential to ensure that these structures remain operational over extended periods. Highway Brazilian bridges were designed based on several codes [1]-[4] over the years, leading to different safety margins between them, highlighted, for instance, by the increase in the vehicular loads that were up to 240kN in [1], 360kN in [2], 450kN in [3] and [4]. In addition, several bridges have been in service for over 40 years exposed to environmental conditions, which can reduce their service life requiring maintenance, repair or rehabilitation in some cases.

The need to evaluate these structures prompted the creation of data inspection systems for existing bridges to provide information about their properties and structural conditions, such as National Bridge Inventory (NBI) [5] in USA and Sistema de Gestão de Obras de Arte Especiais (SGO) [6] in Brazil. In other countries, the agencies responsible for the operation of the bridges published documents with available data, such as Canada [7] and South Korea [8]. These databases are essential to the development of bridge portfolios that include geometric and structural properties of the most representative existing bridges typologies. Bridges’ parameters (e.g., span length, number of spans and column height) may be described by typical or mean values or by probability density functions. In this regard, several studies have performed using reliability analyses by generating samples that are statistically representative of the bridge inventory according to the distribution functions, for instance,  fragility assessment of bridges subjected to seismic loads [8]-[12], hurricane loads [13]-[15], vehicular loads [16], and barge-bridge collision and scour [17]. Such portfolios can also be used to generate typical bridge prototypes, which allows different types of analyses to be performed, such as dynamic and fatigue assessment of bridge-traffic-wind interaction [18]-[20].  

This paper aims to perform a statistical analysis of federal highway bridges located in Northeastern Brazil, in order to develop a representative portfolio. Further information on the motivation behind choosing this region is detailed in the next section. Such bridge portfolio characterization provides valuable information regarding the most representative structural schemes and distributions functions of the parameters necessary to define them properly. Finally, the results of this paper can be used to conduct regional studies on the structural, non-structural and economic studies on bridges, since the individual analysis for each bridge remains a computationally inefficient approach [11].

%%%%%%%%%%%%%%%%%%%%%%%%%%%%%%%%%%%%%%%%
\section{DNIT Management System}
%%%%%%%%%%%%%%%%%%%%%%%%%%%%%%%%%%%%%%%%

In Brazil, the National Department of Infrastructure and Transportation (DNIT) is responsible for supervising the projects, constructions, operations, maintenance, rehabilitation and replacement of more than 5000 bridges on national highways [21]. In order to evaluate these bridges, DNIT developed a management system denominated Sistema de Gestão de Obras de Arte Especiais (SGO), which presents several survey reports that include bridge geometrical characteristics, inspection data, and rating information. From the rating indices, about 45$\%$ of the bridges have minor structural damages and 5$\%$ of them have critical structural damages that require immediate or mid-term interventions [21]. 3306 Brazilian bridges located in federal highways have available information about the year of construction, where 76$\%$ of them are more than 40 years old and 29$\%$ are more than 60 years old [22].

In Northeastern Brazil, there are approximately 2500 bridges located on federal highways under the supervision of DNIT, which represents about 50$\%$ of the total. The region comprises 9 states, an area of approximately 1552167 km², and is the second most populous region in Brazil, with a population of approximately 57.37 million [23]. The bridges in this region present the worst structural condition countrywide; note that 6.6$\%$ of them require immediate or mid-term interventions [21]. Moreover, recurrent floods in the region aggravate bridges' condition, e.g., in the state Pernambuco alone, from 1840 to 2020, at least 11 floods occurred and caused mild to complete damage to bridges [24] such as in the Baeté bridge (PE) that collapsed in 2010. In addition, this region also presents a relatively higher seismicity when compared to other parts of Brazil, as can be observed in [25]-[27]. 

%%%%%%%%%%%%%%%%%%%%%%%%%%%%%%%%%%%%%%%%
\section{Bridge Classification and Characteristics}
%%%%%%%%%%%%%%%%%%%%%%%%%%%%%%%%%%%%%%%%

In order to perform a bridge assessment on a regional scale, it is essential to have an understanding of the characteristics of the existing bridges. Herein, a detailed study is conducted to characterize the bridge inventory in Northeastern Brazil. A group of 250 bridges is randomly collected from the total bridge inventory in order to create a representative bridge portfolio. The number of bridges is limited because of the restricted access to the database. Technical reports in the database include information such as the year of construction, geometric characteristics and number of spans, girders and columns, types of bearings, characteristics of the abutments, among others. From the total amount, concrete bridges represent 92.4$\%$, cast-in-place bridges represent 89.6$\%$, straight bridges represent 94$\%$, and elastomeric bearings are present in 96.7$\%$ of the bridges supported on bearings. Therefore, these are the characteristics initially selected to create the portfolio, since they represent the vast majority of the bridges.

The structural components of the selected bridges are investigated to define the most typical typologies. Based on the analysis of the structural characteristics, bridges are divided in seven types in decreasing order of representativeness (Table \ref{Tab1}) according to the number and continuity of spans, type of abutments and connectivity between the deck and the bents [28]. Non-integral abutments represent the bridges in which the deck is connected to the abutments by elastomeric bearings, and the bridges without abutments are those where transverse beams located at the end of the span contain the landfill.

\begin{table}[h!]
\centering
\scalebox{0.909}{
\begin{tabular}{ccc}
\hline
Bridge type & Bridge description                                                                                                                                                        & $\%$              \\ \hline
\textbf{T1} & \textbf{Single-span and non-integral abutments}                                                                                                                           & \textbf{44.8$\%$} \\ \hdashline
\textbf{T2} & \textbf{\begin{tabular}[c]{@{}c@{}}Multi-span continuous,   connected by elastomeric\\ bearings between the deck and the bents and with\\ no abutments\end{tabular}}      & \textbf{23.2$\%$} \\ \hdashline
\textbf{T3} & \textbf{\begin{tabular}[c]{@{}c@{}}Multi-span continuous,   connected by elastomeric\\ bearings between the deck and the bents and\\ non-integral abutments\end{tabular}} & \textbf{16$\%$}   \\ \hdashline
\textbf{T4} & \textbf{Single-span and integral abutments}                                                                                                                               & \textbf{6$\%$}    \\ \hdashline
T5          & \begin{tabular}[c]{@{}c@{}}Multi-span discontinuous, connected by elastomeric\\ bearings between the deck and the bents and\\ non-integral abutments\end{tabular}         & 4.8$\%$           \\ \hdashline
T6          & \begin{tabular}[c]{@{}c@{}}Multi-span continuous, monolithic connection between\\ the deck and the bents and non-integral abutments\end{tabular}                          & 1.6$\%$           \\ \hdashline
T7          & \begin{tabular}[c]{@{}c@{}}Multi-span continuous, monolithic connection \\ between the deck and the bents and no abutments\end{tabular}                                   & 1.6$\%$           \\ \hdashline
Others      &                                                                                                                                                                           & 2$\%$             \\ \hline
\multicolumn{3}{l}{\textit{Note: Bold font indicates typical bridges}}                                                                                                                                     
\end{tabular}
}
\caption{\small Representativeness and description of the bridges divided into large groups.}
\label{Tab1}
\end{table}

The first four types (T1, T2, T3 and T4) represent 90$\%$ of all selected bridges, so the other types are neglected due to lack of representativeness (less than 5$\%$ of the total). Moreover, the T5 bridges are neglected because there aren’t enough bridges to create reliable statistical distributions able to represent their geometric properties and because the available reports are not complete.

The selected bridge types are divided into classes, according to the deck section and number of columns per bents, as these geometric configurations are expected to impact the structural behavior of the bridges. Three deck sections are common in the inventory (slab deck, T-beam deck and box girder deck), and two bent configurations are representative (wall bent and two column bents). The results described in Table \ref{Tab2} illustrate that four bridge classes (SS-SD-Abut, SS-TB-Abut, MSC-TB and MSC-TB-Abut) that represent 67.6$\%$ (169) of the total of bridges analyzed (250). The same assumption used to discard the bridge types (less than 5$\%$) is used to neglect classes with low representativeness. Nomenclatures are defined only for the classes of bridges analyzed herein.

\begin{table}[h!]
\centering
\scalebox{0.909}{
\begin{tabular}{ccc}
\hline
Bridge class & Bridge description                                                                                                                                                        & $\%$              \\ \hline
\textbf{SS-SD-Abut} & \textbf{\begin{tabular}[c]{@{}c@{}}Single-span, slab deck and non-integral\\  abutments\end{tabular}}                                                                                                                           & \textbf{16.4$\%$} \\ \hdashline
\textbf{SS-TB-Abut} & \textbf{\begin{tabular}[c]{@{}c@{}}Single-span, T-beam deck and non-integral \\ abutments\end{tabular}}      & \textbf{27.6$\%$} \\ \hdashline
\textbf{MSC-TB} & \textbf{\begin{tabular}[c]{@{}c@{}}Multi-span continuous, T-beam deck, two \\ column bents and with no abutments\end{tabular}} & \textbf{17.2$\%$}   \\ \hdashline
- & \begin{tabular}[c]{@{}c@{}}Multi-span continuous, box girder deck, two\\ column bents and with no abutments\end{tabular}                                                                                                                               & 2.8$\%$    \\ \hdashline
\textbf{MSC-TB-Abut}          & \textbf{\begin{tabular}[c]{@{}c@{}}Multi-span continuous, T-beam deck, two \\ column bents and non-integral abutments\end{tabular}}         & \textbf{6.4$\%$}           \\ \hdashline
-          & \begin{tabular}[c]{@{}c@{}}Multi-span continuous, T-beam deck, wall bent \\ and non-integral abutments\end{tabular}                          & 4.0$\%$           \\ \hdashline
-          & \begin{tabular}[c]{@{}c@{}}Multi-span continuous, box girder deck, wall bent \\ and non-integral abutments\end{tabular}                                   & 1.6$\%$           \\ \hdashline
-      &  Single-span, slab deck and integral abutments                                                                                                                                                                         & 4.4$\%$             \\ \hline
\multicolumn{3}{l}{\textit{Note: Bold font indicates typical bridges}}                                                                                                                                     
\end{tabular}
}
\caption{\small Representativeness and description of the bridge classes.}
\label{Tab2}
\end{table}

\begin{figure}[H]
\centering
\centering
\includegraphics[scale=0.60]{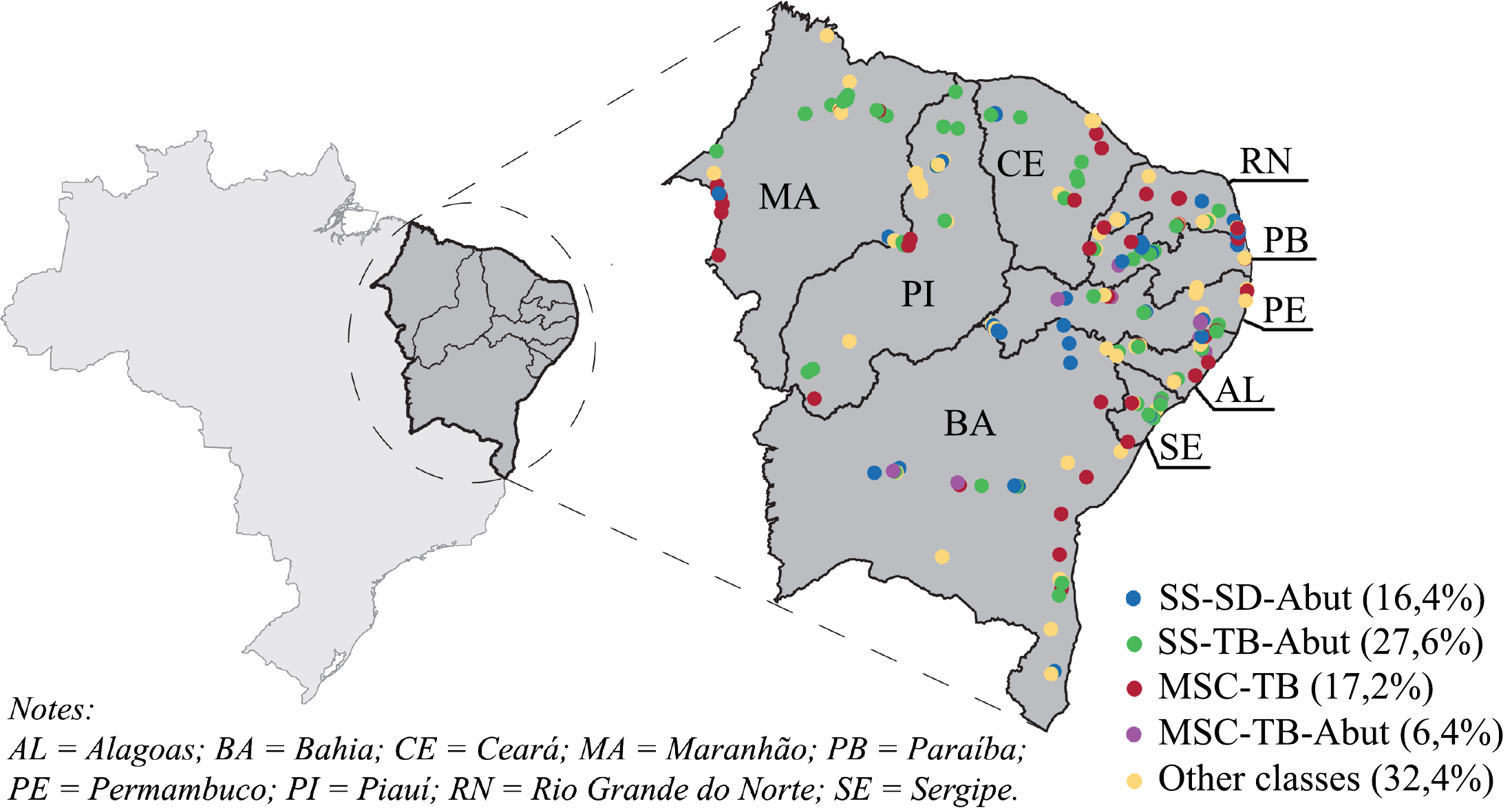}
\caption{\small Location of the bridge classes in Northeastern Brazil.}
\label{Fig1} 
\end{figure}

The location of the selected bridges with their class type is presented in Figure \ref{Fig1}.

\begin{figure}[H]
\centering
\centering
\includegraphics[scale=0.909]{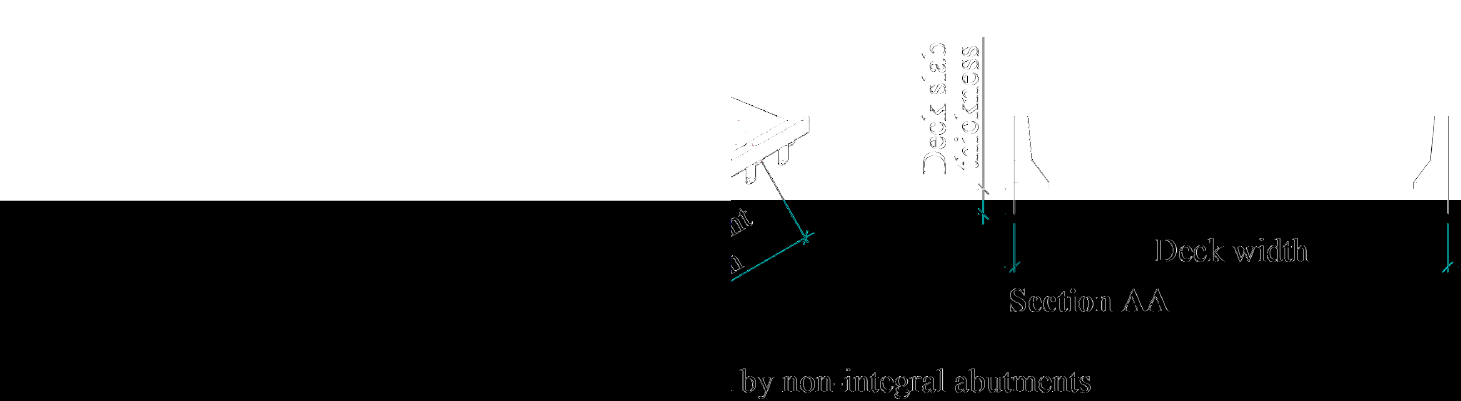}
\caption{\small Geometric properties of the SS-SD-Abut bridge class.}
\label{Fig2} 
\end{figure}

\begin{figure}[H]
\centering
\centering
\includegraphics[scale=0.909]{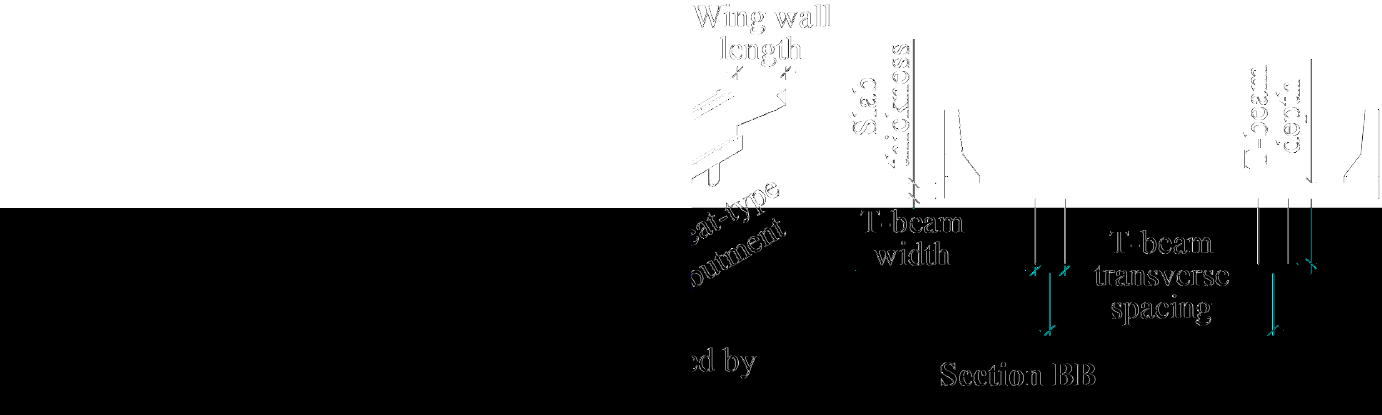}
\caption{\small Geometric properties of the SS-TB-Abut bridge class.}
\label{Fig3} 
\end{figure}

\begin{figure}[H]
\centering
\centering
\includegraphics[scale=0.909]{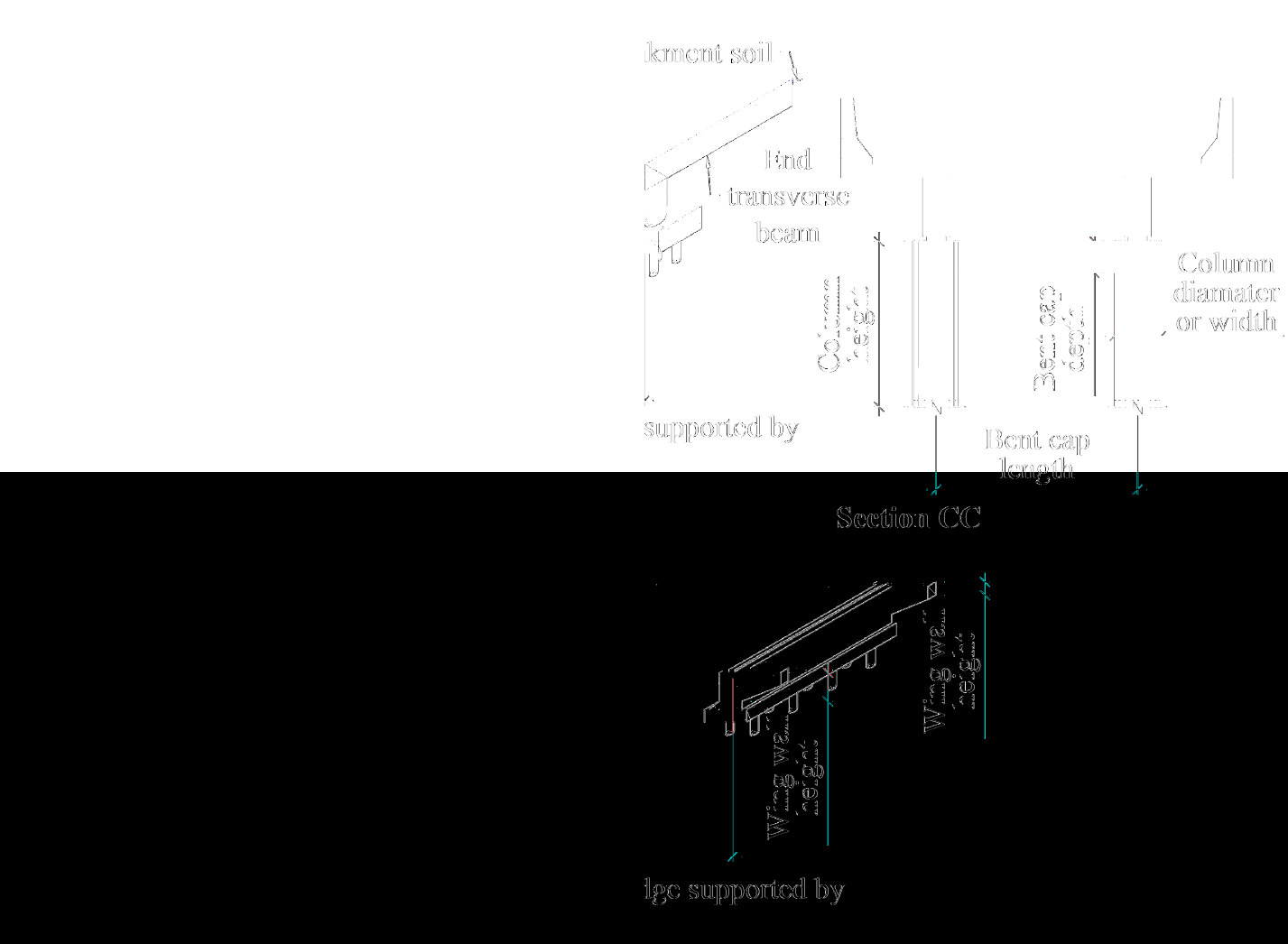}
\caption{\small Geometric properties of the MSC-TB and MSC-TB-Abut bridge classes.}
\label{Fig4} 
\end{figure}

Figures \ref{Fig2}, \ref{Fig3} and \ref{Fig4} illustrate the geometric properties of the typical bridge classes described in Table \ref{Tab1}. Gravity or U seat-type abutments support SS-SD-Abut, SS-TB-Abut and MSC-TB-Abut bridge classes. All decks of the bridge classes (SS-SD-Abut, SS-TB-Abut, MSC-TB and MSC-TB-Abut) are supported on bearings.  The SS-SD-Abut bridge class is simply described as having constant deck slab thickness (Figure \ref{Fig2}); however, this type of bridges tends to have less thickness at the ends of the section to have better aesthetic characteristics and reduce construction costs.

Figure \ref{Fig3} illustrates the SS-TB-Abut bridge class. Therefore, two type of abutments are shown as an illustration; nevertheless, only one type of abutment is found for each bridge. Furthermore, the number of T-beams varies significantly and is considered in the next sections.

Note that the MSC-TB bridge class bridge system (Figure \ref{Fig4}) is not common in many countries; however, several Brazilian bridges adopt the end transverse beams as containment of the embankment soil, which require special treatment in the area to avoid undesirable displacements, such as adopting independent foundation systems for the approach slabs.

%%%%%%%%%%%%%%%%%%%%%%%%%%%%%%%%%%%%%%%%
\section{Characteristics of the Bridge Classes}
%%%%%%%%%%%%%%%%%%%%%%%%%%%%%%%%%%%%%%%%

The next step in the statistical analysis is the study of the most important parameters (i.e., type or absence of bearings, year of construction and live load classes) that influence the geometry characterization of the bridge classes. Some other parameters (i.e., soil characteristics and topography of the site, climate conditions, material properties, reinforcement rates) are epistemic uncertainties [29]-[31], since they are not informed in the reports. Note that the analysis performed in this article uses only the data presented in the survey reports. Further information could be achieved with in-situ investigations, which are unfeasible due to the high costs involved and the amount of bridges. Nevertheless, experimental or statistical studies can be used to define some parameters (i.e., compressive concrete strength, steel yield stress, damping ratio, gap length, elastomeric pads shear modulus) depending on the analysis that will be performed, such as [11], [32]-[35]. The following results are obtained by analyzing the 169 reports used to define the bridge classes, as described in the previous section.

The characterization of the bearing system is essential to predict the structural behavior of the bridges. Of the 169 inspections, only 44 (27$\%$) managed to identify the type of bearing on the bridges, probably due to the deterioration and site conditions or lack of adequate inspection equipment. The geometric and material properties of the bearings are not informed in the reports. Elastomeric bearings are the most commonly adopted in the bridge classes, since they have been identified in 40 out of 44 reports. Therefore, it is assumed that the bearing system is composed of elastomeric bearings in all classes of bridge.

The representativeness of the year of construction of the bridge classes are shown in Table \ref{Tab3}. About 38$\%$ of the SS-TB-Abut bridges are more than 70 years old according to the reports containing this information, meanwhile most of the MSC-TB and MSC-TB-Abut bridge classes were built between 1960 and 1980. The construction year is important to define the design requirements at the time of the bridge construction; however, most of the reports (61$\%$) do not present this information. As a result, any direct relations between the year of construction and the characterization of the geometric parameters of the bridge classes are neglected.

\begin{table}[h!]
\centering
\scalebox{0.909}{
\begin{tabular}{ccccccc}
\hline
Bridge      & \multicolumn{6}{c}{Year of construction ($\%$)}                    \\
class       & 1900-50 & 1951-60 & 1961-1970 & 1971-1980 & 1981-2019 & NI \\ \hline
SS-SD-Abut  & 2.44       & 0.00    & 7.32      & 14.63     & 0.00       & 75.61  \\
SS-TB-Abut  & 15.38      & 7.69    & 9.23      & 6.15      & 1.54       & 60.00  \\
MSC-TB      & 4.76       & 2.38    & 16.67     & 16.67     & 2.38       & 57.14  \\
MSC-TB-Abut & 0.00       & 0.00    & 46.67     & 13.33     & 0.00       & 40.00  \\ \hdashline
All classes & 7.98       & 3.68    & 14.11     & 11.66     & 1.23       & 61.35  \\ \hline
\multicolumn{7}{l}{\textit{Note: NI = Not informed in bridge reports}}
\end{tabular}
}
\caption{\small Representativeness of the year of construction of the bridge classes.}
\label{Tab3}
\end{table}

A similar study is carried out to estimate the representativeness of the live load capacity of the bridge classes, as presented in Table \ref{Tab4}. The live load class should be evaluated, as it influences the design of the structural elements of the bridge. According to the results obtained in the reports, most of the SS-TB-Abut bridges were designed considering live loads of 240 kN. The minority of bridges were designed according to the live loads usually adopted in [4] and most reports (60$\%$) do not inform live load capacities. Therefore, the year of construction and live load capacities are epistemic uncertainties, and can be reduced as more detailed reports become available.

\begin{table}[h!]
\centering
\scalebox{0.909}{
\begin{tabular}{ccccc}
\hline
Bridge      & \multicolumn{4}{c}{Axle load of live loads ($\%$)}                    \\
class       & 240 kN & 360 kN & 450 kN & Not informed \\ \hline
SS-SD-Abut  & 2.44       & 19.51    & 2.44      & 75.61  \\
SS-TB-Abut  & 18.46      & 12.31    & 4.62      & 64.62  \\
MSC-TB      & 11.90       & 23.81    & 16.67     & 47.62  \\
MSC-TB-Abut & 6.67       & 60.00    & 0.00     & 33.33  \\ \hdashline
All classes & 11.66       & 21.47    & 6.75     & 60.12  \\ \hline
\end{tabular}
}
\caption{\small Representativeness of the live load capacity of the bridge classes.}
\label{Tab4}
\end{table}

The structural conditions of the bridge classes are analyzed using the rate indices provided by the reports (Table \ref{Tab5}). Rate indices indicate the level of estimated deterioration conditions in bridges ranging from 1 (severe structural damage) to 5 (no structural damage).

\begin{table}[h!]
\centering
\scalebox{0.909}{
\begin{tabular}{cccccccc}
\hline
Bridge      & Rate  & \multicolumn{6}{c}{Structural elements ($\%$)}      \\
class       & index & Slabs & T-beams & Abutments  & Columns   & EB & Total \\ \hline
            & 1     & 0.00  & -       & 0.00  & -     & -         & 0.00  \\
            & 2     & 4.88  & -       & 2.44  & -     & -         & 7.32  \\
SS-SD-Abut  & 3     & 39.02 & -       & 17.07 & -     & -         & 43.9  \\
            & 4     & 34.15 & -       & 39.02 & -     & -         & 41.46 \\
            & 5     & 21.95 & -       & 41.46 & -     & -         & 7.32  \\ \hdashline
            & 1     & 1.54  & 1.54    & 0.00  & -     & -         & 3.08  \\
            & 2     & 12.31 & 30.77   & 4.62  & -     & -         & 36.92 \\
SS-TB-Abut  & 3     & 27.69 & 16.92   & 15.38 & -     & -         & 16.92 \\
            & 4     & 46.15 & 27.69   & 24.62 & -     & -         & 27.69 \\
            & 5     & 12.31 & 23.08   & 55.38 & -     & -         & 23.08 \\ \hdashline
            & 1     & 0.00  & 0.00    & -     & 2.38  & 0.00      & 2.38  \\
            & 2     & 0.00  & 2.38    & -     & 7.14  & 7.14      & 14.29 \\
MSC-TB      & 3     & 30.95 & 26.19   & -     & 4.76  & 14.29     & 38.1  \\
            & 4     & 54.76 & 52.38   & -     & 35.71 & 50.00     & 42.86 \\
            & 5     & 14.29 & 19.05   & -     & 50.00 & 28.57     & 2.38  \\ \hdashline
            & 1     & 0.00  & 0.00    & 0.00  & 0.00  & -         & 0.00  \\
            & 2     & 0.00  & 6.67    & 0.00  & 6.67  & -         & 6.67  \\
MSC-TB-Abut & 3     & 40.00 & 20.00   & 6.67  & 6.67  & -         & 40.00 \\
            & 4     & 46.67 & 60.00   & 46.67 & 26.67 & -         & 53.33 \\
            & 5     & 13.33 & 13.33   & 46.67 & 60.00 & -         & 0.00  \\ \hline
            
\multicolumn{8}{l}{ \textit{ Note: EB = transverse beams located at the end of the spain that contains } } \\
\multicolumn{8}{l}{ \textit{  the embankment } }
\end{tabular}
}
\caption{\small Representativeness of the rate indices of the structural elements of the bridge classes.}
\label{Tab5}
\end{table}

The SS-TB-Abut bridge class has the highest percentage of bridges (40$\%$) with rating indices 1 and 2, which is consistent with the results of the year of construction (Table \ref{Tab3}) and live load capacities (Table \ref{Tab4}). The MSC-TB bridge class has about 17$\%$ of the bridges with indices of 1 and 2. However, the columns present the worst condition among the other components of the MSC-TB bridge class, which, along with the lack of seismic ductile design, can jeopardize the structural performance of such class. The bridge classes SS-SD-Abut and MSC-TB-Abut are in better condition with more than 90$\%$ of the bridges with a rating of 3 or higher. These indices are not related to the characterization of geometric parameters, but they may be useful to develop models of structural deterioration [36]-[39].

Finally, the year of construction, live load capacities and deterioration indices can be used to estimate the design of the samples generated by the distributions and the functions used to characterize the geometric parameters of the bridge classes described in the next section. 

%%%%%%%%%%%%%%%%%%%%%%%%%%%%%%%%%%%%%%%%
\section{Characterization of Geometric Parameters}
%%%%%%%%%%%%%%%%%%%%%%%%%%%%%%%%%%%%%%%%

Several geometric features are adopted as random variables to define each bridge class: number of spans; span length; slab deck width and thickness; T-beam deck width; depth, width and slab thickness; column height and cross section dimensions; bent cap length, depth and width; abutment type and dimensions. Superstructure dimensions (i.e., span length and T-beam depth) affect stiffness and deck mass. The horizontal load resistance of the bridges depends on the abutment and bent properties and the bearing conditions. The other parameters are correlated, such as bent cap length and T-beam spacing. The foundations, material properties, and design details are not provided in the reports.

\subsection{SS-SD-Abut}

\begin{figure}[H]
\centering

\begin{subfigure}[t]{0.49\textwidth}
\centering
 \includegraphics[scale=0.28]{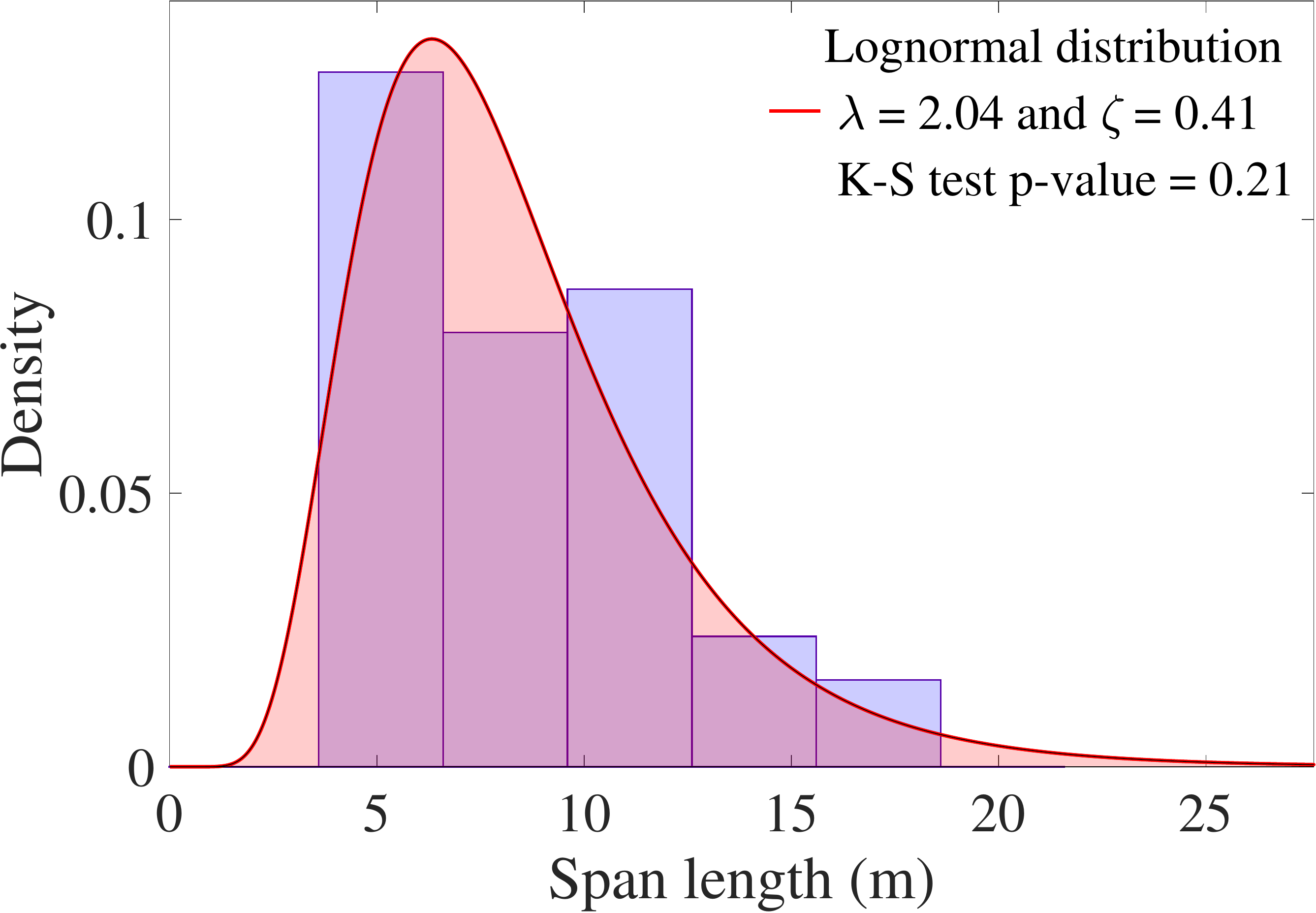}
\caption{\small Span length lognormal distribution} 
\label{Fig5a} 
\end{subfigure} 
\begin{subfigure}[t]{0.49\textwidth}
\centering
\includegraphics[scale=0.28]{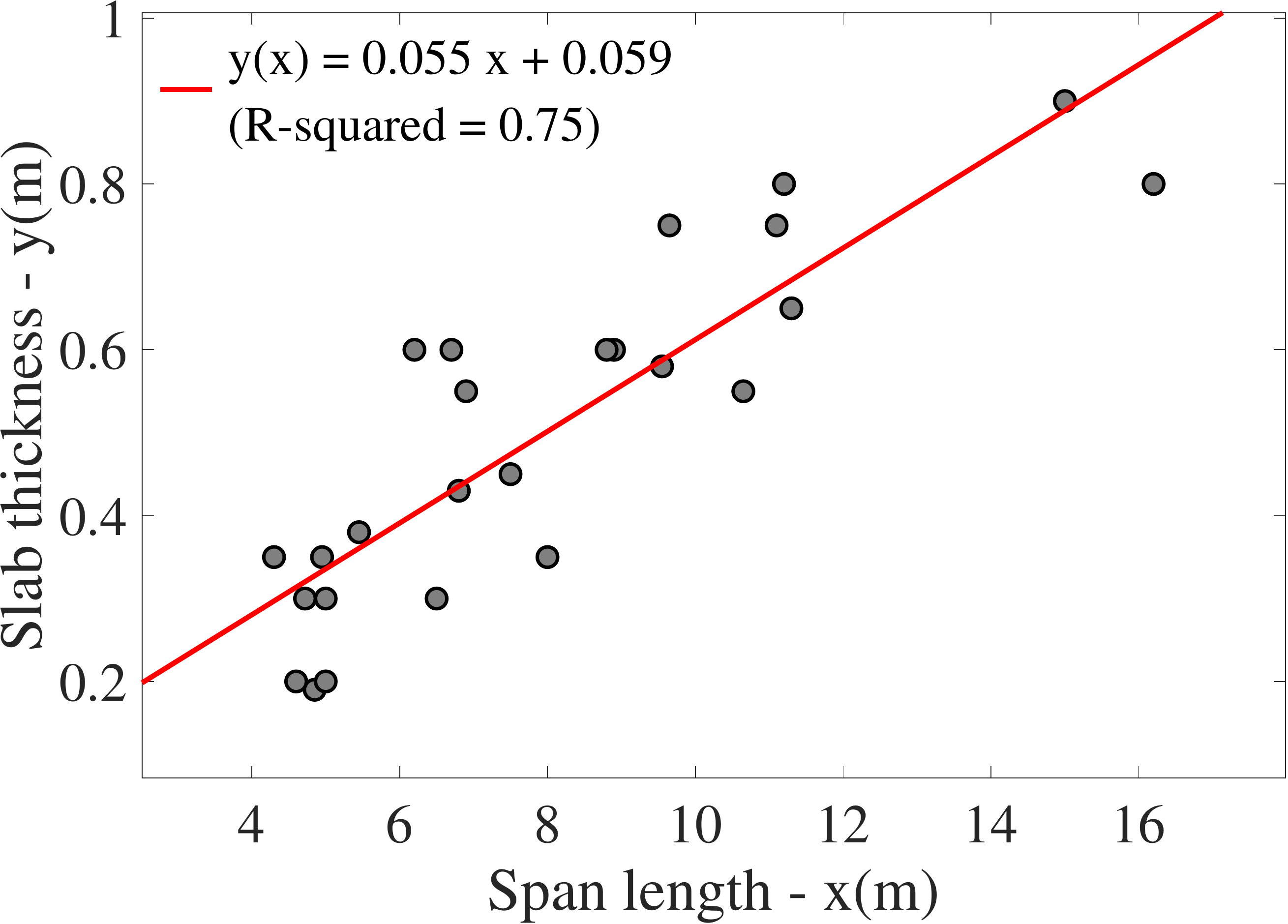}
\caption{\small Deck slab thickness linear regression} 
\label{Fig5b} 
\end{subfigure}

\vspace{0.5cm}

\begin{subfigure}[t]{0.49\textwidth}
\centering
 \includegraphics[scale=0.28]{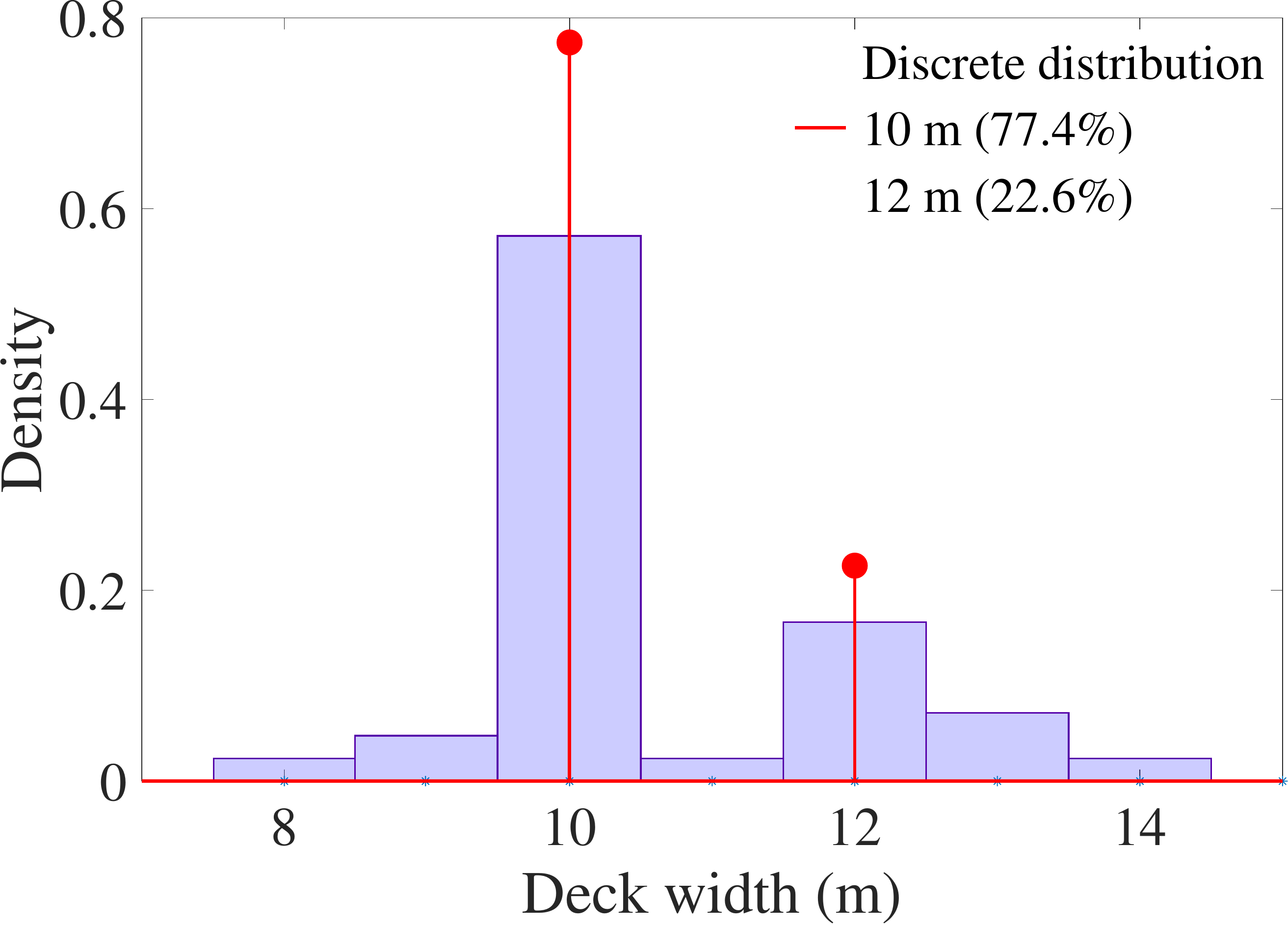}
\caption{\small Deck width discrete distribution} 
\label{Fig5c} 
\end{subfigure} 
\begin{subfigure}[t]{0.49\textwidth}
\centering
\includegraphics[scale=0.28]{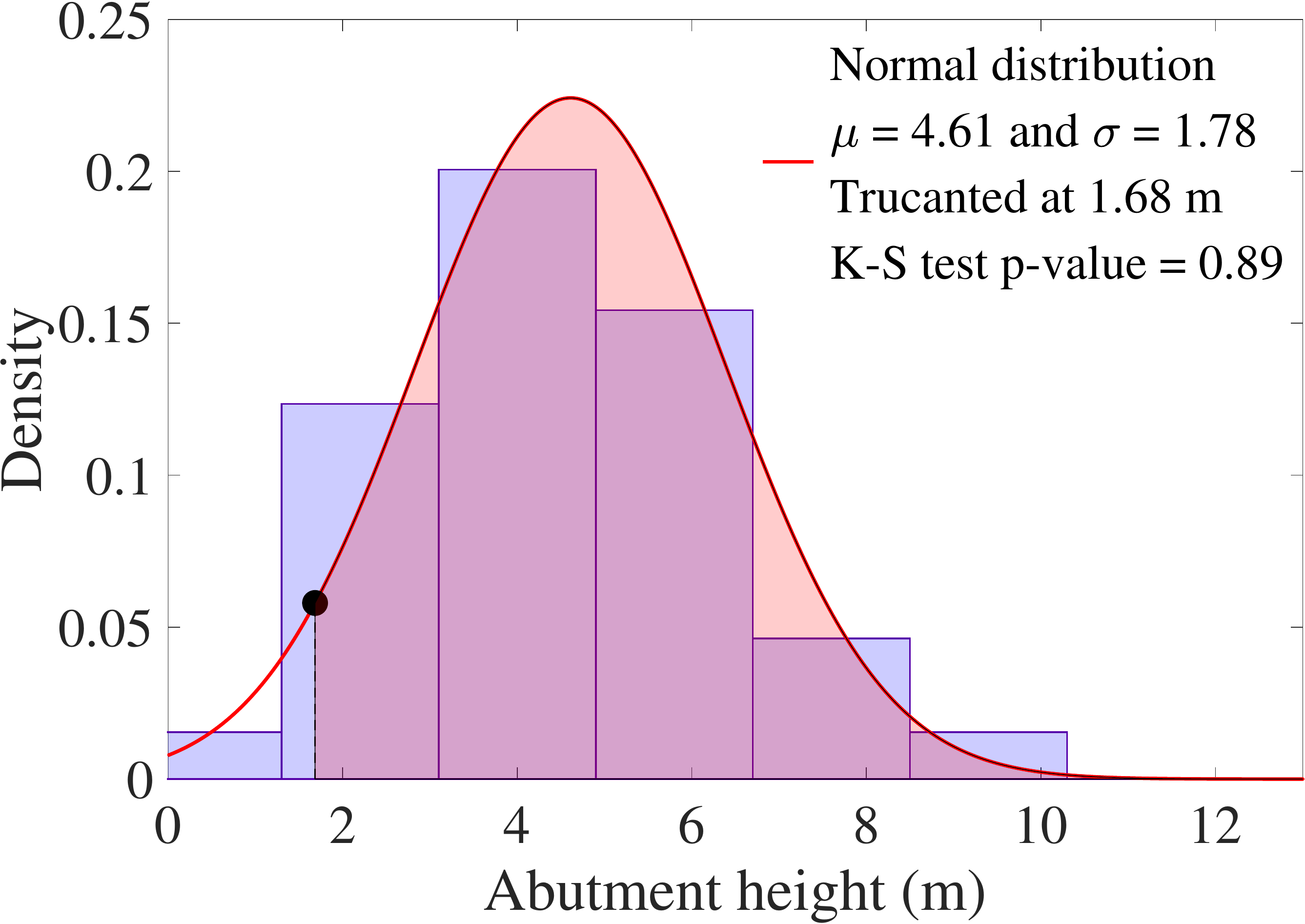}
\caption{\small Abutment height normal distribution} 
\label{Fig5d} 
\end{subfigure}

\caption{\small Bridge parameters of SS-SD-Abut bridge class.}
\label{Fig5} 
\end{figure}

The deck width increases as more lanes become necessary over the years and, depending on the class observed, it changes with the year of construction. Therefore, SS-SD-Abut bridge class presents 10 and 12 meters of deck width, best represented by discrete distributions according to the respective probability of occurrence, as shown in Figure \ref{Fig5}c. 

The span length and abutment height are considered independent and random variables, since both depend on the conditions of the site and the structural scheme (i.e., bridge class). The random variables are described by empirical cumulative distribution functions and then fitted to lognormal or normal distributions (Figure \ref{Fig5}a and \ref{Fig5}d). Kolmogorov-Smirnov (KS) tests and two-sample KS tests are performed to verify which distribution fits each data with a significant cutoff level of 5$\%$, in order to statistically represent the entire population. The lower tail of normal distributions is truncated using a cutoff point based on the 5$\%$ quantile, since these values are compatible with those obtained in the reports (Figure \ref{Fig5}d). This procedure is adopted to avoid negative and unrepresentative physical values.

The slab thickness of the bridge is simply described as functions of the span length through linear regressions (Figure \ref{Fig5}b), since increasing the span length requires a thicker slab to support the loads. This assumption is also adopted in other bridge classes, because of the strong correlation between the two parameters according to Pearson’s linear correlation. In addition, p-values resulting from the test of null hypothesis of no correlation against the  nonzero correlation are generated and a significant cutoff level of 5$\%$ is adopted. Thus, the parameters described by linear regressions have p-values lower than 0.05.

The SS-SD-Abut bridge class has its parameters presented in Table \ref{Tab6}. Note that lognormal and normal distributions are defined by mean ($\lambda$ and $\mu$) and standard deviation ($\zeta$ and $\sigma$). Most of the bridges in this class (92.9$\%$) have gravity seat-type abutments, thus this is defined as representative for the entire class. In all classes, it is assumed that abutment width is equal to deck width.

\begin{table}[h!]
\centering
\scalebox{0.909}{
\begin{tabular}{ccc}
\hline
Variable                    & Distribution or function & Parameters (m)                                                                                        \\ \hline
Span length ($L_s$)         & Lognormal                & $\lambda = 2.04$ and $\zeta = 0.41$                                                                   \\ \hdashline
Deck slab thickness ($S_t$) & Linear regression        & $S_t(L_s) = 0.055 L_s + 0.059$                                                                        \\ \hdashline
Deck width ($D_w$)            & Discrete                 & \begin{tabular}[c]{@{}c@{}}$W_d = 10$ (77.4$\%$)\\ $W_d = 12$ (22.6$\%$)\end{tabular} \\ \hdashline
Abutment height ($H_a$)       & Normal                   & \begin{tabular}[c]{@{}c@{}}$\mu = 4.61$, $\sigma= 1.78$ \\ ($H_a \geq 1.68$)\end{tabular}    \\ \hline         
\end{tabular}
}
\caption{\small Representativeness of the rate indices of the structural elements of the bridge classes.}
\label{Tab6}
\end{table}

\subsection{SS-TB-Abut}

The class is divided into two subclasses according to the abutment type, with its respective probability of occurrence. In all classes with U seat-type abutments, the wing wall heights are considered equal to the T-beam depth and abutment height, due to the lack of data. The heights of U seat-type abutments ($\mu = 5.49$ m) are generally greater than gravity seat-type abutments ($\mu = 4.08$ meters), which is also observed for the results of the MSC-TB-Abut bridge class, as described in Table \ref{Tab10}. The wing wall length is simply considered equal to the abutment height for the SS-TB-Abut and MSC-TB-Abut bridge classes. Table \ref{Tab7} presents the abutment’s parameters for SS-TB-Abut bridge class.

\begin{table}[h!]
\centering
\scalebox{0.909}{
\begin{tabular}{lccc}
\hline
\multicolumn{1}{c}{Variable}        & Distribution & Subclass                                      & Parameters (m)                                                                                                                                \\ \hline
\multicolumn{1}{c}{Abutment height} & Normal       & \begin{tabular}[c]{@{}c@{}}1\\ 2\end{tabular} & \begin{tabular}[c]{@{}c@{}}$\mu = 4.08$, $\sigma = 1.12$ and $H_a \geq 2.25$\\ $\mu = 5.49$, $\sigma = 2.13$ and $H_a \geq 1.99$\end{tabular} \\ \hline
\multicolumn{4}{l}{\textit{Subclass 1 indicates gravity seat-type abutments with 80$\%$ occurrence}}                                                                                                                                        \\
\multicolumn{4}{l}{\textit{Subclass 2 indicates U seat-type abutments with 20$\%$ occurrence}}                                                                                                                                                    
\end{tabular}
}
\caption{\small Parameters of the abutments for SS-TB-Abut bridge class.}
\label{Tab7}
\end{table}

The parameters for the deck and span for SS-TB-Abut bridge class are described in Table \ref{Tab8}. They are divided into three subclasses according to the number of T-beams, with its respective probabilities of occurrence. Bridges with more than four T-beams are discarded due to lack of representativeness (less than 10$\%$). The span length data is divided according to the number of T-beams, since an increase in the number of T-beams tends to impact the dimensions of the deck. Therefore, the distributions and functions of the deck variables are defined for each subclass. The T-beam depth values are adjusted by linear regression depending on the span length, as required in structural design.

\begin{table}[h!]
\centering
\scalebox{0.909}{
\begin{tabular}{cccc}
\hline
Variable                                                                                                                                              & \begin{tabular}[c]{@{}c@{}}Distribution\\ or function\end{tabular}           & Subclass           & Parameters (m)                                         \\ \hline
\multirow{3}{*}{\begin{tabular}[c]{@{}c@{}}Span length\\ ($L_s$)\end{tabular}}                                                                        & \multirow{3}{*}{Lognormal}                                                   & 1                  & $\lambda = 2.46$ and $\zeta = 0.29$                    \\
                                                                                                                                                      &                                                                              & 2                  & $\lambda = 1.95$ and $\zeta = 0.34$                    \\
                                                                                                                                                      &                                                                              & 3                  & $\lambda = 2.63$ and $\zeta = 0.37$                    \\ \hdashline
\multirow{3}{*}{\begin{tabular}[c]{@{}c@{}}T-beam depth\\ ($D_b$)\end{tabular}}                                                                       & \multirow{3}{*}{\begin{tabular}[c]{@{}c@{}}Linear\\ regression\end{tabular}} & 1                  & $D_b(L_s) = 0.09 L_s + 0.23$                          \\
                                                                                                                                                      &                                                                              & 2                  & $D_b(L_s) = 0.03 L_s + 0.49$                          \\
                                                                                                                                                      &                                                                              & 3                  & $D_b(L_s) = 0.07 L_s + 0.05$                          \\ \hdashline
\multirow{6}{*}{\begin{tabular}[c]{@{}c@{}}Deck width ($D_w$),\\ T-beam transverse\\ spacing ($T_{bs}$)\\ and slab thickness ($S_t$)\end{tabular}} & \multirow{6}{*}{Discrete}                                                    & \multirow{2}{*}{1}                  & $D_w = 8$, $T_{bs} = 5.8$ and $S_t = 0.24$ (61.1$\%$)  \\
                                                                                                                                                      &                                                                              &                    & $D_w = 10$, $T_{bs} = 4.8$ and $S_t = 0.27$ (38.9$\%$) \\ \cdashline{3-4}
                                                                                                                                                      &                                                                              & \multirow{2}{*}{2} & $D_w = 8$, $T_{bs} = 3.1$ and $S_t = 0.24$ (82.4$\%$)  \\
                                                                                                                                                      &                                                                              &                    & $D_w = 12$, $T_{bs} = 3.0$ and $S_t = 0.28$ (17.6$\%$) \\ \cdashline{3-4} 
                                                                                                                                                      &                                                                              & \multirow{2}{*}{3} & $D_w = 8$, $T_{bs} = 1.8$ and $S_t = 0.32$ (37.5$\%$)  \\
                                                                                                                                                      &                                                                              &                    & $D_w = 10$, $T_{bs} = 2.5$ and $S_t = 0.27$ (62.5$\%$) \\ \hline
\multicolumn{4}{l}{\textit{Subclass 1 indicates a deck section with two T-beams with 40.4$\%$ occurrence}}                                                                                                                                                                                                         \\
\multicolumn{4}{l}{\textit{Subclass 2 indicates a deck section with three T-beams with 40.4$\%$ occurrence}}                                                                                                                                                                                                       \\
\multicolumn{4}{l}{\textit{Subclass 3 indicates a deck section with four T-beams with 19.2$\%$ occurrence}}                                                                                                                                                                                                       
\end{tabular}
}
\caption{\small Deck and span parameters for the SS-TB-Abut bridge class.}
\label{Tab8}
\end{table}

In addition, as the year of construction affects the number of lanes and deck width, the deck geometry are significantly different, and, thus, deck width, T-beam transverse spacing and slab thickness are not described as statistical distributions for this class. In order to properly define these parameters, groups are created with mean values for each deck width range and its respective probabilities of occurrence, as represented in Table \ref{Tab8}. As an example, 61.1$\%$ of the bridges with two T-beams have the mean values of deck width, T-beam transverse spacing and slab thickness equal to 8, 5.8 and 0.24 m, respectively.

\subsection{MSC-TB}

The deck cross section of all MSC-TB bridges is composed of two T-beams, as illustrated in Figure \ref{Fig4}. The deck width is related to the bent cap length through linear regression, since the bent cap length is equal to the internal slab width. In addition, the number of spans is described by discrete distribution [29] and [30], with 51.4$\%$ of the bridges having three spans, 11.4$\%$ four spans and 37.1$\%$ five spans. The values of intermediate span lengths are assumed to be random variables fitted by normal distribution. In this class, it is common to adopt the length of the end span as a function of the length of the intermediate span to have a better structural behavior [40], as depicted in Table \ref{Tab9}. 

\begin{figure}[H]
\centering

\begin{subfigure}[t]{0.49\textwidth}
\centering
 \includegraphics[scale=0.28]{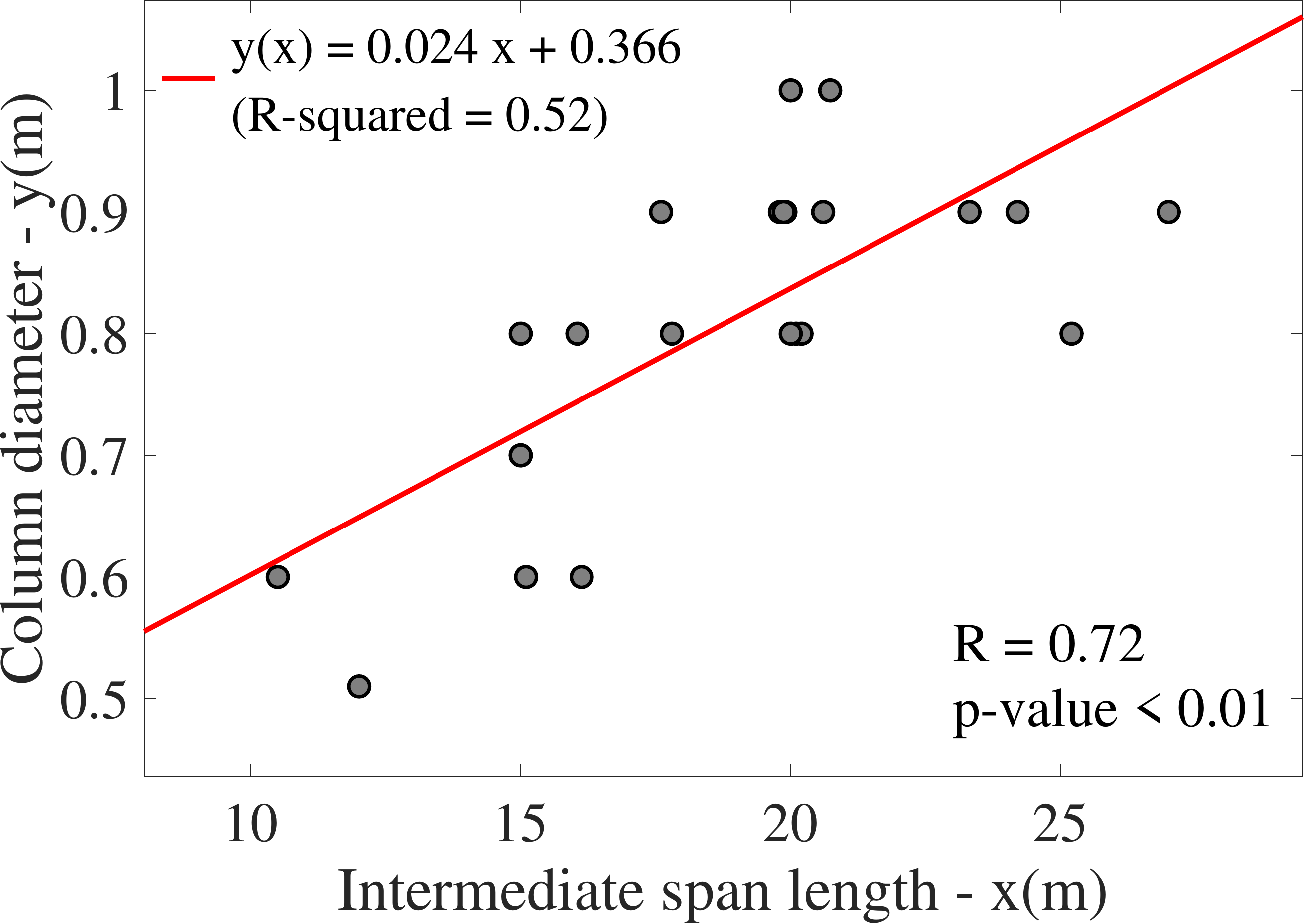}
\caption{\small Correlation of span length and \\ column diameter} 
\label{Fig6a} 
\end{subfigure} 
\begin{subfigure}[t]{0.49\textwidth}
\centering
\includegraphics[scale=0.28]{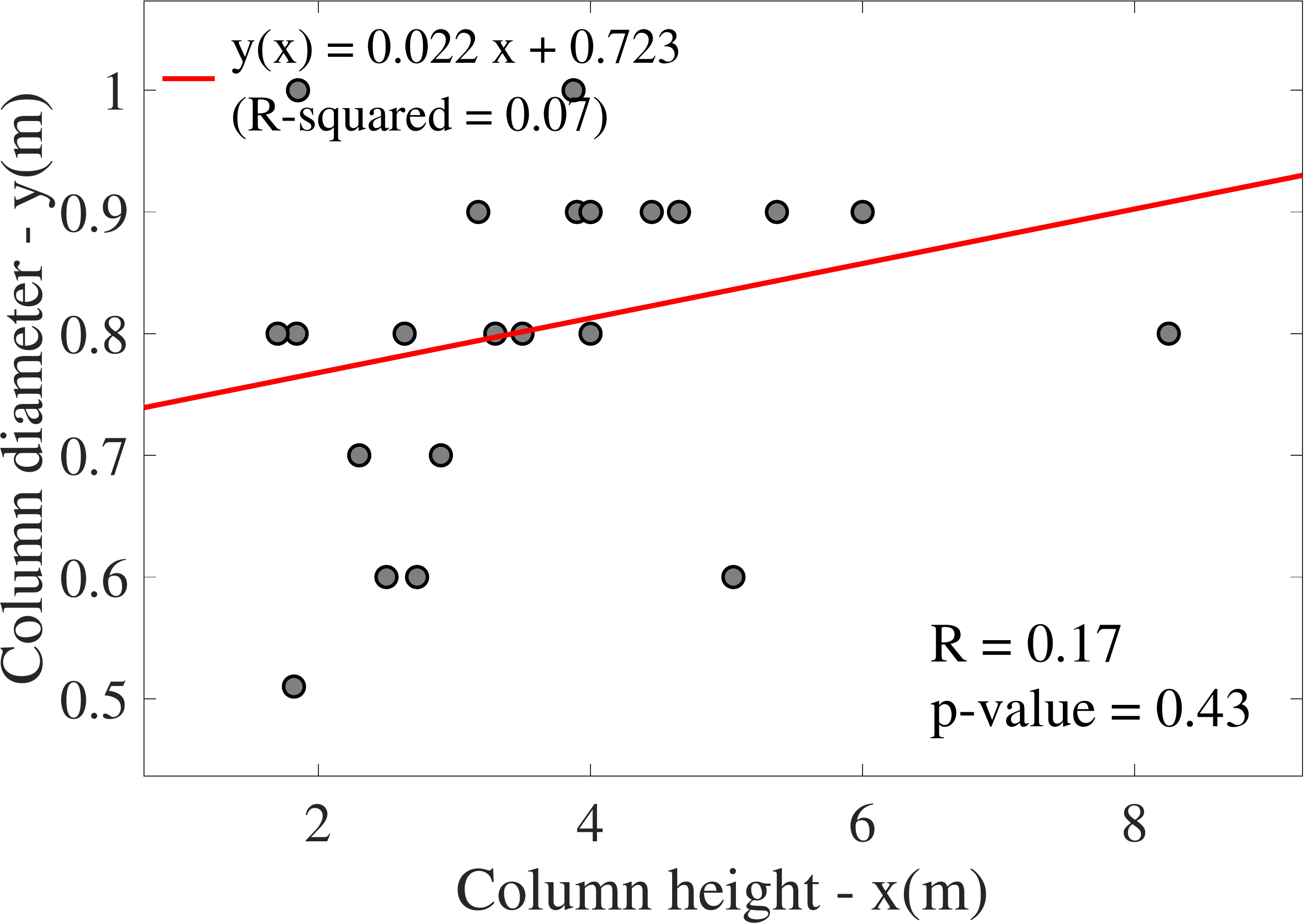}
\caption{\small Correlation of deck width and \\ column diameter} 
\label{Fig6b} 
\end{subfigure}

\vspace{0.5cm}

\begin{subfigure}[t]{1\textwidth}
\centering
\includegraphics[scale=0.28]{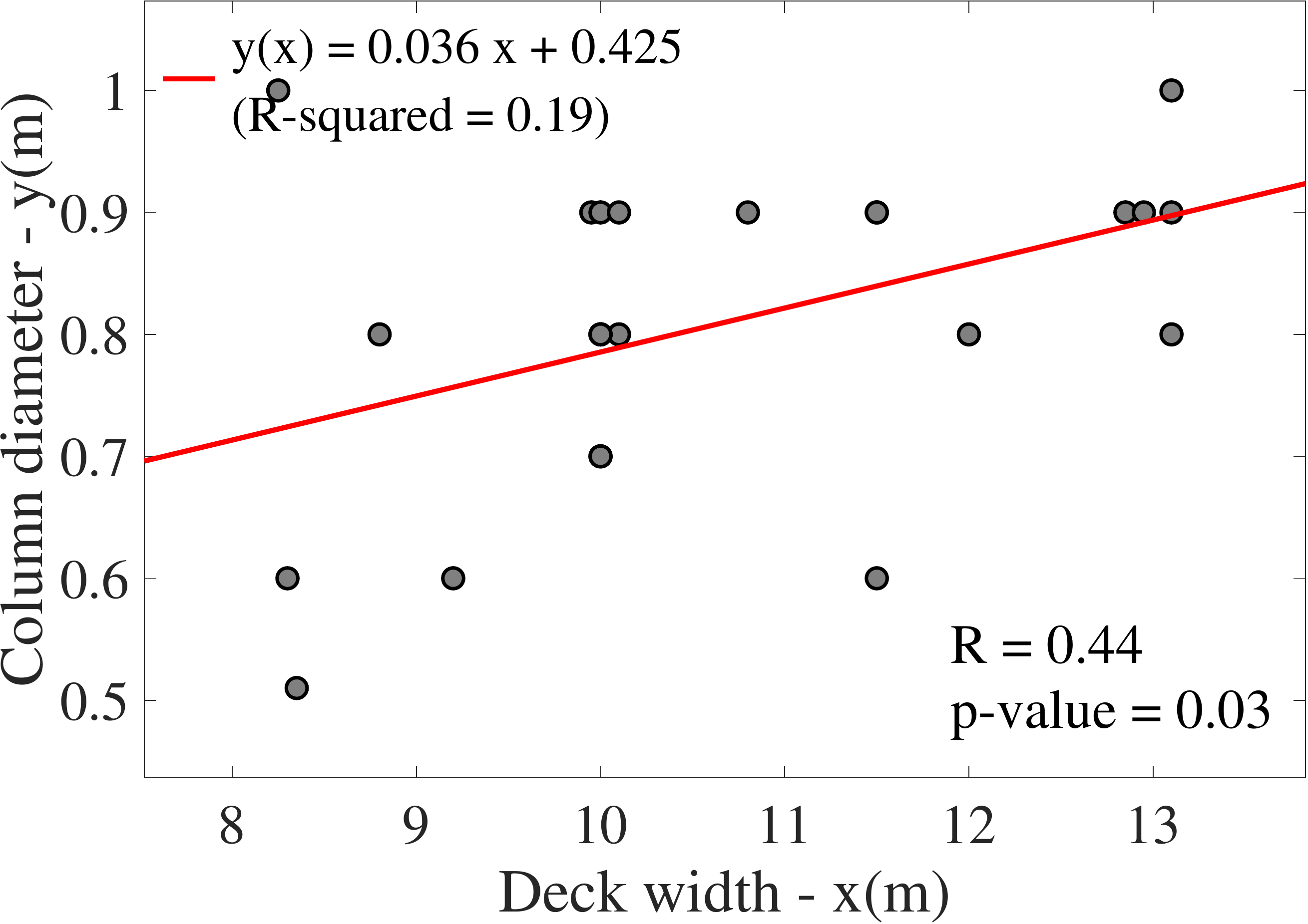}
\caption{\small Correlation of Column height and column diameter} 
\label{Fig6c} 
\end{subfigure}

\caption{\small Correlation of column dimensions for the MSC-TB Bridge class.}
\label{Fig6} 
\end{figure}

The MSC-TB bridge class is divided into two subclasses according to the geometry of the column cross section. Circular columns represent 73.7$\%$ of the total bridges in this class, while rectangular columns represent 26.3$\%$. The increase in the span length and the deck width tends to generate bridges with greater masses that lead to greater dimensions of the columns. The column height also affects the design of the columns due to the bending moments generated by horizontal forces (i.e., braking and wind loads). The correlation between these parameters is studied to properly define the column dimensions (Figure \ref{Fig6}). Pearson’s linear correlation coefficients (R) indicate that the strongest correlation is between column diameter and span length. Therefore, it is considered that column height and column diameter are not correlated, since the calculated p-value is 0.43. Finally, the column dimensions are only described as functions of the span length to simplify the analysis, since there is no strong correlation between the diameter and the deck width. The values of the column heights are considered as random variables fitted by lognormal distribution independent of the column cross section, since the height of the bridge depends on the characteristics of the site where the bridge is implemented.

Table \ref{Tab9} presents the bridge parameters for MSC-TB class. The values of length and depth of the bent cap are adopted as linear regressions, since the bent cap length tends to be designed according to the deck width to reduce the slab thickness due to bending moments. For this reason, the values of the slab thickness are adjusted by linear regression according to the values of the bent cap length. Only a few reports present T-beams and bent caps width and, for that reason, the mean value of 0.4 m is adopted.

\begin{table}[h!]
\centering
\scalebox{0.909}{
\begin{tabular}{cccc}
\hline
Variable                                                                                      & \begin{tabular}[c]{@{}c@{}}Distribution\\ or function\end{tabular} & Subclass                                      & Parameters (m)                                                                                                          \\ \hline
Number of spans ($N_s$)                                                                       & Discrete                                                           & -                                             & \begin{tabular}[c]{@{}c@{}}$N_s = 3$ (51.4$\%$)\\ $N_s = 4$ (11.4$\%$)\\ $N_s = 5$ (37.1$\%$)\end{tabular}              \\ \hdashline
\begin{tabular}[c]{@{}c@{}}Intermediate span \\ length ($L_{is}$)\end{tabular}                & Normal                                                             & -                                             & \begin{tabular}[c]{@{}c@{}}$\mu = 18.24$, $\sigma = 4.79$\\ and $L_{is} \geq 10.4$\end{tabular}                         \\ \hdashline
End span length ($L_{es}$)                                                                    & Linear regression                                                  & -                                             & $L_{es}(L_{is}) = 0.18 L_{is} + 1.03$                                                                                    \\ \hdashline
T-beam depth ($D_{b}$)                                                                    & Linear regression                                                  & -                                             & $D_{b}(L_{is}) = 0.05 L_{is} + 0.78$                                                                                    \\ \hdashline
Deck width ($D_w$)                                                                            & Lognormal                                                          & -                                             & $\lambda = 2.35$ and $\zeta = 0.14$                                                                                     \\ \hdashline
Bent cap length ($L_{bc}$)                                                                    & Linear regression                                                  & -                                             & $L_{bc} (D_w) = 0.56 D_w + 0.64$                                                                                        \\ \hdashline
Bent cap depth ($D_{bc}$)                                                                     & Linear  regression                                                 & -                                             & $D_{bc} (L_{bc}) = 0.12 L_{bc} + 0.33$                                                                                  \\ \hdashline
Slab thickness ($S_d$)                                                                        & Linear regression                                                  & -                                             & $S_d(L_{bc}) = 0.04 L_{bc}$                                                                                             \\ \hdashline
Column height ($H_c$)                                                                         & Lognormal                                                          & -                                             & $\lambda = 1.25$ and $\zeta = 0.47$                                                                                     \\ \hdashline
\begin{tabular}[c]{@{}c@{}}Column diameter ($\phi_c$)\\ and Column width ($D_c$)\end{tabular} & Linear regression                                                  & \begin{tabular}[c]{@{}c@{}}1\\ 2\end{tabular} & \begin{tabular}[c]{@{}c@{}}$\phi_c (L_{is}) = 0.02 L_{is} + 0.37$\\ $D_c (L_{is}) = 0.03 L_{is} + 0.25$\end{tabular} \\ \hline
\multicolumn{4}{l}{\textit{Subclass 1 indicates bridges composed of circular columns with 73.7$\%$ occurrence}}                                                                                                                                                                                                                             \\
\multicolumn{4}{l}{\textit{Subclass 2 indicates bridges composed of rectangular columns with 26.3$\%$ occurrence}}                                                                                                                                                                                                                         
\end{tabular}
}
\caption{\small Deck and span parameters for the MSC-TB bridge class.}
\label{Tab9}
\end{table}

\subsection{MSC-TB-Abut}

The parameters of MSC-TB-Abut bridge class are illustrated in Table \ref{Tab10}. All reports pointed out that the thickness of the slabs is 0.3 m, and the width and depth of the bent caps are 0.4 m and 1 m, respectively. Therefore, those values are assumed constant for those parameters. The deck width adopted is 10 m, since it represents 80$\%$ of the bridges. The deck section is composed of two T-beams, since it represents 93.3$\%$ of all bridges. 

\begin{figure}[h!]
\centering

\begin{subfigure}[t]{0.49\textwidth}
\centering
 \includegraphics[scale=0.28]{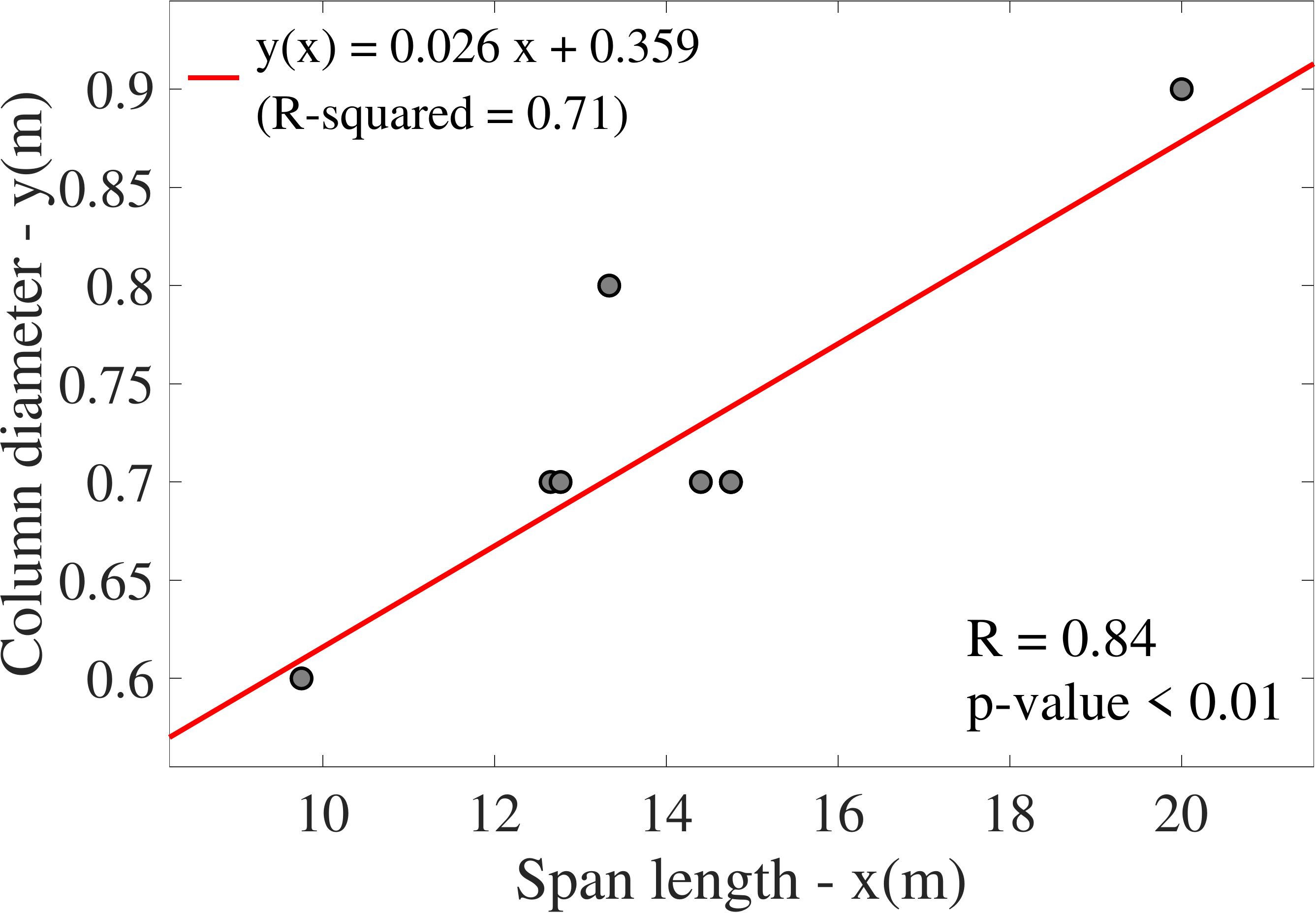}
\caption{\small Correlation of span length and \\ column diameter} 
\label{Fig7a} 
\end{subfigure} 
\begin{subfigure}[t]{0.49\textwidth}
\centering
\includegraphics[scale=0.28]{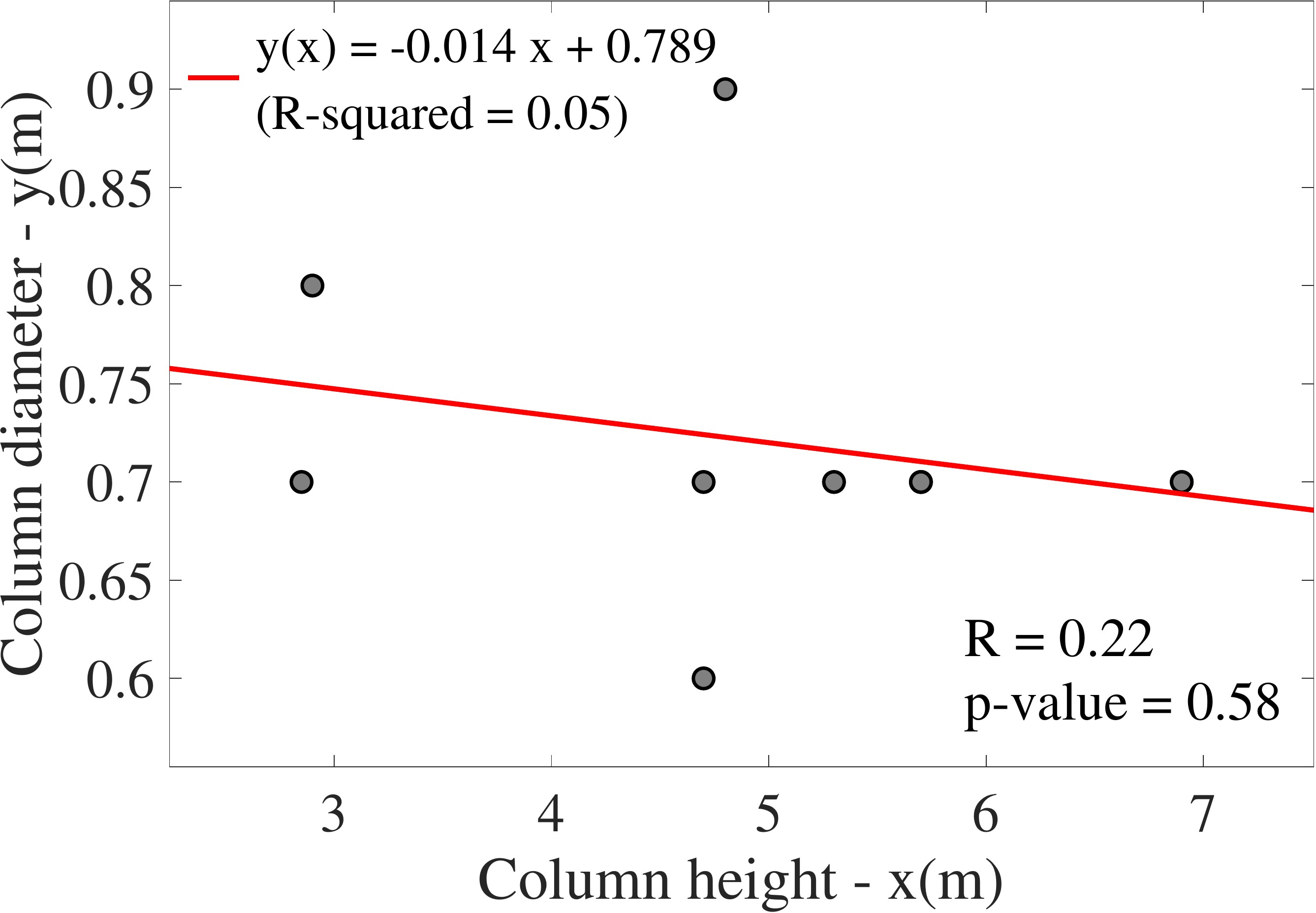}
\caption{\small Correlation of Column height and \\ column diameter} 
\label{Fig7b} 
\end{subfigure}

\caption{\small Correlation of column dimensions for the MSC-TB Bridge class.}
\label{Fig7} 
\end{figure}

All reports for this bridge class indicate that the columns have circular cross sections. The same approach used in the previous bridge class is adopted herein to verify there is a correlation between the diameter of the columns and the length of the spans (Figure \ref{Fig7}a) and the height of the columns (Figure \ref{Fig7}b). Pearson’s linear correlation coefficient (0.84) and the p-value (<0.01) of the correlation between the diameter of the columns and the length of the spans suggests a stronger correlation than those obtained with the correlation between the diameter and the height of the columns (R=-0.22 and p-value=0.58), as illustrated in Figure \ref{Fig7}. Therefore, the diameter of the columns of this class are related only to the length of the spans by linear regressions (Table \ref{Tab10}).

The number of spans are described as discrete functions and the span length follows a lognormal distribution, where the lengths of each span is considered equal to simplify the analysis. The assumptions for T-beam depth are the same as for the previous bridge class. However, the bent cap length is considered a random variable fitted by lognormal distribution, since the deck width has a constant value according to the reports. Note that the standard deviation value is very small (0.04) and the bent cap length could be assumed to be a constant value of 4.8 m. Two subclasses are created, according to the abutment type. The abutment height values follow a lognormal distribution that depends on the type of abutment.

\begin{table}[h!]
\centering
\scalebox{0.909}{
\begin{tabular}{cccc}
\hline
Variable                   & \begin{tabular}[c]{@{}c@{}}Distribution\\ or function\end{tabular} & Subclass                                      & Parameters (m)                                                                                                    \\ \hline
Number of spans ($N_s$)    & Discrete                                                           & -                                             & \begin{tabular}[c]{@{}c@{}}$N_s = 2$ (66.7$\%$)\\ $N_s = 3$ (26.7$\%$)\\ $N_s = 4$ (6.6$\%$)\end{tabular}         \\ \hdashline
Span length ($L_{s}$)      & Normal                                                             & -                                             & $\lambda = 2.7$ and $\zeta = 0.37$                                                                                \\ \hdashline
T-beam depth ($D_{b}$)     & Linear regression                                                  & -                                             & $D_b (L_{s}) = 0.09 L_{s} + 0.07$                                                                                  \\ \hdashline
Bent cap length ($L_{bc}$) & Lognormal                                                          & -                                             & $\lambda = 1.57$ and $\zeta = 0.04$                                                                               \\ \hdashline
Column diameter ($\phi_c$) & Linear  regression                                                 & -                                             & $\phi_c (L_{s}) = 0.03 L_{s} + 0.36$                                                                             \\ \hdashline
Column height ($H_c$)      & Normal                                                             & -                                             & \begin{tabular}[c]{@{}c@{}}$\mu = 4.82$, $\sigma = 1.58$\\ and $H_c \geq 2.22$\end{tabular}                       \\ \hdashline
Abutment height ($H_a$)    & Lognormal                                                          & \begin{tabular}[c]{@{}c@{}}1\\ 2\end{tabular} & \begin{tabular}[c]{@{}c@{}}$\lambda = 1.65$ and $\zeta = 0.41$\\ $\lambda = 1.96$ and $\zeta = 0.13$\end{tabular} \\ \hline
\multicolumn{4}{l}{\textit{Subclass 1 indicates gravity seat-type abutments with 42.9$\%$ occurrence}}                                                                                                                                                              \\
\multicolumn{4}{l}{\textit{Subclass 2 indicates U seat-type abutments with 57.1$\%$ occurrence}}                                                                                                                                                                   
\end{tabular}
}
\caption{\small Deck and span parameters for the MSC-TB bridge class.}
\label{Tab10}
\end{table}

More details on the characterization of these geometric parameters can be seen elsewhere [41].

%%%%%%%%%%%%%%%%%%%%%%%%%%%%%%%%%%%%%%%%
\section{Conclusions}
%%%%%%%%%%%%%%%%%%%%%%%%%%%%%%%%%%%%%%%%

This paper presents a statistical analysis of typical bridges in Northeastern Brazil. Information from DNIT reports (i.e., span length, column height, abutment height and deck width) is collected from 250 bridges to describe the structural properties and geometric characteristics of each group of bridges separated into classes. The bridges are divided into four classes representing 169 (67.6$\%$) of the 250 bridges (100$\%$). The first and second classes represent single-span bridges supported by non-integral abutments with different deck configurations (slab and T-beam); multi-span continuous bridges unsupported by abutments represent the third class; multi-span continuous bridges supported by non-integral abutments are grouped in the fourth class.

Geometric parameters are described by distribution functions as uncertain independent variables (i.e., span length and column height) or by linear regressions as dependent variables (i.e., T-beam depth as a function of span length). The type of abutment, the number of T-beams and the geometry of the column cross section are separated into subclasses to adequately describe dependent variables (i.e., abutment height, T-beam depth and column diameter) with their respective probability density. These results allow the generation of statistical representative geometric samples, which can be complemented by new in-situ investigations, new reports or published data depending on the type of analysis that will be performed.

This systematic statistical characterization of the Northeastern Brazil bridge portfolio provide insights on typical parameters and design details, and can pave the path to future regional scale analyses related to bridge safety and resilience quantification. For example,  future bridge assessment studies should use the geometries of this bridge portfolio to evaluate regional impacts due to extreme events (i.e., floods, earthquakes, fires, hurricanes and explosion), updates to current codes (i.e. increasing live loads and consideration of seismic events) and effects of aging on the structure (i.e., fatigue and corrosion).

%%%%%%%%%%%%%%%%%%%%%%%%%%%%%%%%%%%%%%%%
\section*{Acknowledgements}
%%%%%%%%%%%%%%%%%%%%%%%%%%%%%%%%%%%%%%%%

This study was financed in part by the Coordenação de Aperfeiçoamento de Pessoal de Nível Superior - Brasil (CAPES) - Finance Code 001; and by the São Paulo Research Foundation (FAPESP) - Finance Code 2018/23304-9. The opinions, findings, and conclusions or recommendations expressed in this paper are those of the authors only and do not necessarily reflect the views of the sponsors or affiliates.

%%%%%%%%%%%%%%%%%%%%%%%%%%%%%%%%%%%%%%%%
\section*{References}
%%%%%%%%%%%%%%%%%%%%%%%%%%%%%%%%%%%%%%%%

\begin{flushleft}

[1] Associação Brasileira de Normas Técnicas \textbf{NB 6: Carga móvel em pontes de concreto armado}. Rio de Janeiro, RJ, Brasil: ABNT, 1943.

[2] Associação Brasileira de Normas Técnicas \textbf{NB 6: Carga móvel em pontes de concreto armado}. Rio de Janeiro, RJ, Brasil: ABNT, 1960.

[3] Associação Brasileira de Normas Técnicas \textbf{NBR 7188: Carga móvel em ponte rodoviária e passarela de pedestres}. Rio de Janeiro, RJ, Brasil: ABNT, 1982.

[4] Associação Brasileira de Normas Técnicas \textbf{NBR 7188: Road and pedestrian live load on bridges, viaducts, footbridges and other structures}. Rio de Janeiro, RJ, Brasil: ABNT, 2013.

[5] Federal Highway Administration, \textbf{National Bridge inventory}. Washington, DC, 2021.

[6] Departamento Nacional de Infraestrutura de Transportes, \textbf{Sistema de Gerenciamento de Obras de Arte Especiais}. Brasil, 2021.

[7] Ministère des Transports du Québec, \textbf{Manuel d’inspection des structures: évaluation des dommages}. Canadá, 2017.

[8] S. M. Lee and T. J. Kang, “Development of fragility curves for bridges in Korea,” KSCE \textbf{Journal of Civil Engineering}, vol. 11, no. 3, pp. 165-174, 2007, https://doi.org/10.1007/BF02823897.

[9] B. G. Nielson and R. DesRoches, “Analytical seismic fragility curves for typical bridges in the central and southeastern United States,” \textbf{Earthquake Spectra}, vol. 23, no. 3, pp. 615-633, 2007, https://doi.org/10.1193/1.2756815.
 
[10] I. F. Moschonas, A. J. Kappos, P. Panetsos, V. Papadopoulos, T. Makarios and P. Thanopoulos, “Seismic fragility curves for greek bridges: methodology and case studies,” \textbf{Bulletin of Earthquake Engineering}, vol. 7, no. 2, pp. 439-468, 2009, https://doi.org/10.1007/s10518-008-9077-2.

[11] D. H. Tavares, J. E. Padgett and P. Paultre, “Fragility curves of typical as-built highway bridges in eastern Canada,” \textbf{Engineering Structures}, vol. 40, pp. 107-118, 2012,  https://doi.org/10.1016/j.engstruct.2012.02.019.

[12] O. Avsar and A. Yakut, “Seismic vulnerability assessment criteria for rc ordinary highway bridges in Turkey,” \textbf{Structural Engineering and Mechanics}, vol. 43, no. 1, pp. 127-145, 2012, https://doi.org/10.12989/sem.2012.43.1.127.

[13] N. Ataei  and J. E. Padgett, “Probabilistic modeling of bridge deck unseating during hurricane events,” \textbf{Journal of Bridge Engineering}, vol. 18, no. 4, pp. 275-286, 2013, https://doi.org/10.1061/(ASCE)BE.1943-5592.0000371.

[14] N. Ataei  and J. E. Padgett, “Fragility surrogate models for coastal bridges in hurricane prone zones,” \textbf{Engineering Structures}, vol. 103, pp. 203-213, 2015, https://dx.doi.org/10.1016/j.engstruct.2015.07.002.

[15] G. P. Balomenos, S. Kameshwar  and J. E. Padgett, “Parameterized fragility models for multi-bridge classes subjected to hurricane loads,” \textbf{Engineering Structures}, vol. 208, pp. 110213, 2020, https://doi.org/10.1016/j.engstruct.2020.110213.

[16] S. Kameshwar  and J. E. Padgett, “Parameterized fragility assessment of bridges subjected to pier scour and vehicular loads,” \textbf{Journal of Bridge Engineering}, vol. 23, no. 7, pp. 0401844, 2018, https://doi.org/10.1061/(ASCE)BE.1943-5592.0001240.

[17] S. Kameshwar  and J. E. Padgett, “Response and fragility assessment of bridge columns subjected to barge-bridge collision and scour ,” \textbf{Engineering Structures}, vol. 168, pp. 308-319, 2018, https://doi.org/10.1016/j.engstruct.2018.04.082.

[18] C. S. Cai and S. R. Chen, “Framework of vehicle-bridge-wind dynamic analysis,” \textbf{Journal of Wind Engineering and Industrial Aerodynamics}, vol. 92, no. 7-8, pp. 579-607, 2004, https://doi.org/10.1016/j.jweia.2004.03.007.

[19] S. R. Chen and J. Wu, “Modeling stochastic live load for long-span bridge based on microscopic traffic flow simulation,” \textbf{Computers $\&$ structures}, vol. 89, no. 9-10, pp. 813-824, 2011, https://doi.org/10.1016/j.compstruc.2010.12.017.

[20] Y. Zhou and S. Chen, \textbf{Dynamic assessment of bridge deck performance considering realistic bridge-traffic interaction}. Mountain Plains Consortium, Colorado, United States, 2017.

[21] C. B. L. Oliveira, M. Greco and T. N. Bittencourt, “Analysis of the Brazilian federal bridge inventory,” \textbf{IBRACON Struct. Mater. J.}, vol. 12, no. 1, pp. 1-13, 2019, https://dx.doi.org/10.1590/s1983-41952019000100002.

[22] P. T. C. Mendes, “\textbf{Contributions for the Brazilian road network concrete bridge management model},” Ph.D. thesis, Departamento de Engenharia, Universidade de São Paulo, USP, São Paulo, 2009.

[23] Instituto Brasileiro de Geografia e Estatística, \textbf{Censo Brasileiro de 2010}. Rio de Janeiro, 2010.

[24] F. H. B. Alves, “\textbf{Advanced flood forecasting: analysis and improvements of rain forecast, hydrological and hydroghanic models},” M.S. thesis, Universidade Federal de Pernambuco, UFPE, Pernambuco, 2017.

[25] D. Giardini, P. Basham and M. Bery, “The global seismic hazard assessment program (GSHAP) – 1992/1999,” \textbf{Istituto Nazionale di Geofisica e Vulcanologia}, vol. 42, no. 6, pp. 957-974, 1999, https://doi.org/10.4401/ag-3780.

[26] M. Assumpção, M. Pirchiner, J. Dourado and L. Barros, “Terremotos no Brasil: preparando-se para eventos raros,” \textbf{Boletim SBGf}, no. 96, pp. 25-29, 2016.

[27] M. D, Petersen et al., “Seismic hazard, risk, and design for South America,” \textbf{Bulletin of the Seismological Society of America}, vol. 108, no. 2, pp. 781-800, 2018, https://doi.org/10.1785/0120170002.

[28] Federal Emergency Management Agency, \textbf{HAZUS-MH MR1: Technical Manual. Vol. Earthquake model}. Washington, DC, 2003.

[29] B. G. Nielson, “\textbf{Analytical fragility curves for highway bridges in moderate seismic zones}” Ph.D. thesis, Georgia Institute of Technology, Georgia, CA, USA, 2005.

[30] O. Avsar, “\textbf{Fragility based seismic vulnerability assessment of ordinary highway bridges in Turkey}” Ph.D. thesis, Middle East Technical University, Ancara, Turkey, 2009.

[31] D. H. Tavares, “\textbf{Seismic risk assessment of bridges in Quebec using fragility curves}” Ph.D. thesis, Sherbrooke University, Sherbrooke, Canadá, 2012.

[32] G. H. Siqueira, A. S. Sanda, P. Paultre and J. E. Padgett, “Fragility curves for isolated bridges in eastern Canada using experimental results,” \textbf{Engineering Structures}, vol. 74, pp. 311-324, 2014, https://dx.doi.org/10.1016/j.engstruct.2014.04.053.

[33] W. Santiago and A. Beck, “A new study of Brazilian concrete strength conformance,” \textbf{IBRACON Struct. Mater. J.}, vol. 10, no. 4, pp. 906-923, 2017, https://dx.doi.org/10.1590/s1983-41952017000400008.

[34] C. G. Nogueira, “\textbf{Desenvolvimento de modelos mecânicos de confiabilidade e de otimização para aplicação em estruturas de concreto armado},” Ph.D. thesis, Departamento de Engenharia, Universidade de São Paulo, USP, São Paulo, 2010.

[35] S. A. Mirza and J. G. MacGregor, “Variability of mechanical properties of reinforcing bars,” \textbf{Journal of Structural Division}, vol. 105, no. 5, pp. 921-937, 1979.

[36] J. Ghosh and J. E. Padgett, “Aging considerations in the development of time-dependent seismic fragility curves,” \textbf{Journal of Structural Engineering}, vol. 136, no. 12, pp. 1497-1511, 2010, https://doi.org/10.1061/(asce)st.1943-541x.0000260.

[37] S. J. W. Bush, T. F. P. Henning, A. Reith and J. M. Ingham, “Development of a bridge deterioration model in a data-constrained environment,” \textbf{Journal of Performance of Constructed Facilities}, vol. 31, no. 5, pp. 04017080, 2017, https://doi.org/10.1061/(asce)cf.1943-5509.0001074.

[38] H. Li, L. Li, G. Zhou and L. Xiu, “Effects of various modeling uncertainty parameters on the seismic response and seismic fragility estimates of the aging highway bridges,” \textbf{Bulletin of Earthquake Engineering}, vol. 18, no. 14, pp. 6337-6373, 2020, https://doi.org/10.1007/s10518-020-00934-9.

[39] F. Cui, H. Li, X. Dong, B. Wang, J. Li, H. Xue and M. Qi, “Improved time-dependent seismic fragility estimates for deteriorating RC bridge substructures exposed to chloride attack,” \textbf{Advances in Structural Engineering}, vol. 24, no. 3, pp. 437-452, 2021, https://doi.org/10.1177/1369433220956812.

[40] G. H. Siqueira, D. H. Tavares, P. Paultre and J. E. Padgett, “Performance evaluation of natural rubber seismic isolators as a retrofit measure for typical multi-span concrete bridges in eastern Canada,” \textbf{Engineering Structures}, vol. 74, pp. 300-310, 2014, https://dx.doi.org/10.1016/j.engstruct.2014.03.009.

[41] G. H. F. Cavalcante, \textbf{Regional bridge inventory assessment}, 1th ed. Self-published using CreateSpace Independent Publishing Platform, 2021.

\end{flushleft}

\end{document}